\documentclass{llncs}
\usepackage{makeidx}  
\usepackage[english]{babel}
\usepackage[latin1]{inputenc}
\usepackage[T1]{fontenc}
\usepackage{amssymb}
\usepackage{amsmath}
\usepackage{stmaryrd}
\usepackage{mathrsfs}
\usepackage{epsfig}
\usepackage{color}
\usepackage{url}
\usepackage{amsmath}
\usepackage{subfigure}
\usepackage{array}

\newcommand\SetFigFont[5]{}

\newcommand{\Sim}[1]{{\bf Sim}}

\newtheorem{thm}{Theorem}
\newtheorem{prop}{Proposition}

\begin{document}

\setlength\abovecaptionskip{0ex}
\setlength\belowcaptionskip{0ex}

\title{Adversary lower bounds for nonadaptive quantum algorithms}

\author{Pascal Koiran\inst{1} \and  Jürgen Landes\inst{2}
\and Natacha Portier\inst{1} \and Penghui Yao\inst{3}}
\institute{LIP\footnote{UMR 5668 ENS Lyon, CNRS, UCBL associée à l'INRIA.
Work done when Landes and Yao were visiting LIP with financial support from the Mathlogaps program.}, 
Ecole Normale Supérieure de Lyon, Université de Lyon
\and
School of Mathematics, University of Manchester
\and State Key Laboratory of Computer Science, Chinese Academy of Sciences}

\maketitle

\abstract{We present general methods for proving 
lower bounds on the query complexity of nonadaptive quantum algorithms.
Our results are based on the adversary method of Ambainis.}

\section{Introduction}

In this paper we present general methods for proving 
lower bounds on the query complexity of nonadaptive quantum algorithms.
A nonadaptive algorithm makes all its queries simultaneously.
By contrast, an unrestricted (adaptive) algorithm  may choose
its next query based on the results of previous queries.
In classical computing, classes of problems for which adaptivity does not
help have been identified~\cite{BYKS01,KNPproba} and it is known that this
question is connected to a longstanding open problem~\cite{Ros73} 
(see~\cite{KNPproba} for a more extensive discussion).
In quantum computing, the study of nonadaptive algorithms seems especially 
relevant since some of the best known quantum algorithms (namely, Simon's algorithms and some other hidden subgroup algorithms) 
are nonadaptive.
This is nevertheless a rather understudied subject in quantum computing.

The paper that is most closely related to the present work is~\cite{NY04}
(and~\cite{GJ04} is another related paper). 
In~\cite{NY04} the authors use an ``algorithmic argument'' (this is a kind
of Kolmogorov argument) to give lower bounds on the nonadaptive quantum 
query complexity of ordered search, and of generalizations of this problem.
The model of computation that they consider is less general than ours (more on this in section~\ref{model}).

The two methods that have proved most successful in the quest for quantum lower
bounds are the polynomial method 
(see for instance~\cite{BBCMW01,AS04,KNPicalp05,KNP07}) and the adversary
method of Ambainis.
It is not clear how the polynomial method might take the nonadaptivity 
of algorithms into account.
Our results are therefore based on the adversary method, 
in its weighted version~\cite{A06}.
We provide two general lower bounds which yield optimal results for a number of
problems: search in an ordered or unordered list,
element distinctness, graph connectivity or bipartiteness.
To obtain our first lower bound we treat the list of queries performed 
by a nonadaptive algorithm as one single ``super query''.
We can then apply the adversary method to this 1-query algorithm.
Interestingly, the lower bound that we obtain is very closely related to
the lower bounds on {\em adaptive} probabilistic query complexity due to 
Aaronson~\cite{Aaronson04}, and to Laplante and Magniez~\cite{LM04}.
Our second lower bound requires a detour through the so-called minimax (dual)
method and is based on the fact that in a nonadaptive algorithm, 
the probability of performing any given
query is independent of the input.

\section{Definition of the Model} \label{model}

In the black box model, an algorithm accesses its input by querying
a function $x$ (the {\em black box}) from a finite set $\Gamma$ to a 
(usually finite) set~$\Sigma$. 
At the end of the computation, the algorithm decides to accept or reject~$x$,
or more generally produces an output in a (usually finite) set~$S'$.
The goal of the algorithm is therefore to compute a (partial) 
function $F : S \rightarrow S'$, where 
$S=\Sigma^{\Gamma}$ is the set of black boxes.
For example, in the \emph{Unordered Search} problem
$\Gamma=[N]=\{1,\ldots,N\}$, $\Sigma=\{0,1\}$ and
$F$ is the OR function:
$\displaystyle F(x)=\bigvee\limits_{1\leq i \leq N} x(i).$

Our second example is \emph{Ordered Search}.
The sets $\Gamma$ and $\Sigma$ are as in the first example, but $F$
is now a partial function: we assume that the black box satisfies 
the promise that there exists an index $i$
such that $x(j)=1$ for all $j \geq i$, and $x(j)=0$ for all $j<i$.
Given such an $x$, the algorithm tries to compute $F(x)=i$.

A 
quantum algorithm $\mathcal{A}$ that makes $T$ queries
 can be formally described as a tuple $(U_0, \dots, U_T)$,
where each $U_i$ is a unitary operator. For $x\in S$ we define the
unitary operator $O_x$ (the ``call to the black box'') by
$O_x|i\rangle |\varphi\rangle| \psi\rangle=|i\rangle|\varphi\oplus
x(i)\rangle|\psi\rangle$. The algorithm 
$\mathcal{A}$ computes the final state $U_T O_x U_{T-1} \dots U_1 O_x
U_0|0\rangle$ and makes a measurement of some of its qubits. 
The result of this measure
is by definition the outcome of the computation of $\mathcal{A}$ on input~$x$. For a given $\varepsilon$, the query complexity of
a function $F$, denoted $Q_{2,\varepsilon}$, is the smallest query
complexity of a quantum algorithm computing $F$ with probability
of error at most $\varepsilon$.

In the sequel, the quantum algorithms as described above will also be called ad{\em adaptive} to distinguish them from nonadaptive quantum algorithms.
Such an algorithm performs all its 
queries at the same time. 
A nonadaptive black-box quantum algorithm $\mathcal{A}$
that makes $T$ queries can therefore be defined by
a pair  $(U,V)$ of unitary operators.  
For $x\in S$ we define
the unitary operator $O_x^T$ by
$$O_x^T|i_1, \dots, i_T\rangle|\varphi_1, \dots, \varphi_T\rangle|\psi\rangle= 
|i_1, \dots, i_T\rangle|\varphi_1\oplus x(i_1), \dots, \varphi_T\oplus   
x(i_T)\rangle |\psi\rangle.$$
The algorithm $\mathcal{A}$ computes the final state $V O_x^T U|0\rangle$ 
and makes a measurement of some of its qubits.
As in the adaptive case, the result of this measure
is by definition the outcome of the computation of $\mathcal{A}$ on input~$x$.
For a given $\varepsilon$, the nonadaptive query complexity of
a function $F$, denoted $Q^{na}_{2,\varepsilon}$, is the smallest query
complexity of a nonadaptive quantum algorithm computing $F$ with probability
of error at most $\varepsilon$.
Our model is more general than the model of~\cite{NY04}.
In that model, the $|\varphi\rangle$ register must remain set to 0 after application 
of $U$. After application of $O_x^T$, the content of this register is 
therefore equal to $|x(i_1), \dots, x(i_T)\rangle$ rather than $|\varphi_1\oplus x(i_1), \dots, \varphi_T\oplus x(i_T)\rangle$.

It is easy to verify that for every nonadaptive quantum algorithm
$\mathcal{A}$ of query complexity $T$ there is an adaptive quantum algorithm
$\mathcal{A}'$ that makes the same number of queries and computes the same
function, so that $Q_{2,\varepsilon}\leq
Q^{na}_{2,\varepsilon}$. Indeed, consider for every $k\in[T]$ the unitary operator
$A_k$ which maps the state $|i_1, \dots, i_T\rangle|\varphi_1, \dots,
\varphi_T\rangle$
to $$|i_k\rangle |\varphi_k\rangle |i_1,
\dots, i_{k-1}, i_{k+1}, \dots 
i_T\rangle |\varphi_1, \dots, \varphi_{k-1}, \varphi_{k+1}, \dots,
\varphi_T\rangle.$$
If the nonadaptive algorithm $\mathcal{A}$ is defined by the pair 
of unitary operators $(U,V)$,  then the
adaptive algorithm $\mathcal{A}'$ defined by the tuple of unitary operators
$$(U_0, \dots, U_T)=(A_1 U, A_2 A_1^{-1}, \dots, A_T A_{T-1}^{T-1}, V A_T^{-1})$$
computes
the same function.

\section{A Direct Method}

\subsection{Lower Bound Theorem and Applications}

The main result of this section is Theorem~\ref{na-qwm}.
It yields an optimal $\Omega(N)$ lower bound on the nonadaptive quantum
query complexity of Unordered Search and Element Distinctness.
First we 
recall the weighted adversary method of Ambainis and some related definitions.
The constant $C_{\varepsilon}=(1-2\sqrt{\varepsilon(1-\varepsilon)})/2$
will be used throughout the paper.
\begin{definition}\label{weight-function}
The function $w: S^2 \rightarrow R_+$ is a 
\emph{\bf valid weight function} if 
every pair $(x,y)\in S^2$ is assigned a non-negative weight
$w(x,y)=w(y,x)$ that satisfies $w(x,y)=0$ whenever $F(x)=F(y)$.
We then define for all $x\in S$ and $i\in \Gamma$:
$wt(x)=\sum_y w(x,y)$ and
$v(x,i)=\sum_{y:\ x(i)\neq y(i)} \  w(x,y)$.
\end{definition}

\begin{definition}\label{weight-scheme}
The pair $(w,\ w')$ is a \emph{\bf valid weight scheme} if:
\begin{itemize}
\item Every pair $(x,y)\in S^2$ is assigned a non-negative weight
$w(x,y)=w(y,x)$ that satisfies $w(x,y)=0$ whenever $F(x)=F(y)$.
\item Every triple $(x,y,i)\in S^2 \times \Gamma$ is assigned a
non-negative weight $w'(x,y,i)$ that satisfies $w'(x,y,i)=0$ whenever
$x(i)=y(i)$ or $F(x)=F(y)$, and $w'(x,y,i)w'(y,x,i)\geq w^2(x,y)$ for
all $x, y, i$ with $x(i)\neq y(i)$.
\end{itemize}
We then define for all $x\in S$ and $i\in \Gamma$ 
$wt(x)=\sum_y w(x,y)$ and 
$v(x,i)=\sum_y w'(x,y,i)$.
\end{definition}

Of course these definitions are relative to the partial function
$F$. 

\begin{remark}
Let $w$ be a valid weight function and define $w'$ such that 
if $x(i)\neq y(i)$ then $w'(x,y,i)=w(x,y)$ and  $w'(x,y,i)=0$ otherwise. 
Then $(w,w')$ is a valid weight scheme and the functions $wt$ and $v$
defined for $w$ in Definition~\ref{weight-function} are exactly those
defined for $(w,w')$ in Definition~\ref{weight-scheme}.
\end{remark}

\begin{thm}[weighted adversary method of Ambainis~\cite{A06}]
\label{qwm}
Given a probability of error $\varepsilon$ and a partial function $F$,
the quantum query complexity $Q_{2,\varepsilon}(F)$ of $F$ as defined in section~\ref{model} satisfies:

$$Q_{2,\varepsilon}(F) \geq C_{\varepsilon} \max_{(w,w') \ \text{valid}} 
\ \min_{\substack{x,y,i\\ w(x,y)> 0\\ x(i)\neq y(i)}} 
\sqrt{ \frac{wt(x)wt(y)}{v(x,i)v(y,i)} }.$$

\end{thm}


A probabilistic version of this lower bound theorem was obtained by Aaronson~\cite{Aaronson04} and by Laplante and Magniez~\cite{LM04}.
\begin{thm}
\label{pwm}
Fix the probability of error to $\varepsilon=1/3$.
The probabilistic query complexity $P_2(F)$ of $F$ satisfies the lower bound
$P_{2}(F) = \Omega(L_P(F))$, where $$L_P(F)=
\max_w \min_{\substack{x,y,i\\ w(x,y)> 0\\ x(i)\neq y(i)}} \max
\left( 
\frac{wt(x)}{v(x,i)}, \frac{wt(y)}{v(y,i)}    \right).$$
Here $w$ ranges over the set of valid weight functions.
\end{thm}

We now state the main result of this section.
\begin{thm}[nonadaptive quantum lower bound, direct method]
\label{na-qwm}
The nonadaptive query complexity $Q^{na}_{2,\varepsilon}(F)$
of $F$ satisfies the lower bound
$Q^{na}_{2,\varepsilon}(F) \geq C_{\varepsilon}^2 L_Q^{na}(F)$,
where $$L_Q^{na}(F)=
\max_w \max_{s\in S'} \min_{\substack{x,i\\ F(x)=s}}
\frac{wt(x)}{v(x,i)}.$$
Here $w$ ranges over the set of valid weight functions.
\end{thm}
The following theorem, which is an unweighted adversary method for
nonadaptive algorithm, is a consequence of Theorem~\ref{na-qwm}.

\begin{thm} \label{uw}
Let $F: \Sigma^{\Gamma} \rightarrow \{0;1\}$, $X\subseteq F^{-1}(0)$,
$Y\subseteq F^{-1}(1)$ and let $R\subset X \times Y$ be a relation such that:
\begin{itemize}
\item for every $x\in X$ there are at least $m$ elements $y\in Y$ such
that $(x,y)\in R$,
\item for every $y\in Y$ there are at least $m'$ elements $x\in X$ such
that $(x,y)\in R$,
\item for every $x\in X$ and every $i\in \Gamma$ there
are at most $l$ elements $y\in Y$ such that $(x,y)\in R$ and $x(i)\neq
y(i)$, 
\item for every $y\in X$ and every $i\in \Gamma$ there
are at most $l'$ elements $x\in X$ such that $(x,y)\in R$ and $x(i)\neq
y(i)$.
\end{itemize}
Then
$\displaystyle Q^{na}_{2,\varepsilon}(F)\geq C_{\varepsilon}^2 \max(\frac{m}{l}, \frac{m'}{l'}).$
\end{thm}

\begin{proof} As in~\cite{A06} and~\cite{LM04} we set $w(x,y)=w(y,x)=1$ for all $(x,y)\in R$.
Then $wt(x)\geq m$ for all
$x\in A$, $wt(y)\geq m'$ for all $y\in B$, $v(x,i) \leq l$ and
$v(y,i)\leq l'$. \hfill $\boxempty$
\end{proof}
For the Unordered Search problem defined in Section~\ref{model} we have $m=N$ and $l=l'=m'=1$. Theorem~\ref{uw} therefore yields an optimal $\Omega(N)$ 
lower bound. 
The same bound can be obtained for the Element Distinctness problem.
Here the set $X$ of negative instances 
is made up of all one-to-one functions $x:[N] \rightarrow [N]$ and $Y$
contains the functions  $y:[N] \rightarrow [N]$ that are not one-to-one.
We consider the relation $R$ such that $(x,y) \in R$ if and only if 
there is a unique $i$ such that $x(i) \neq y(i)$.
Then $m=2,l=1,m'=N(N-1)$ and $l'=N-1$. 

As pointed out in~\cite{LM04}, the $\Omega(\max(m/l,m'/l'))$ lower bound
from Theorem~\ref{uw} is also a lower bound on $P_2(F)$.
There is a further connection:
\begin{proposition} \label{comparison}
For any function $F$ we have $L_P(F) \geq L_Q^{na}(F)$. 
That is, ignoring constant 
factors, the lower bound on $P_2(F)$ given by Theorem~\ref{pwm} is at least as
high as the lower bound on $Q^{na}_{2,\varepsilon}(F)$ given by Theorem~\ref{na-qwm}.
\end{proposition}
\begin{proof}
Pick a weight function $w_Q$ which is optimal for the ``direct method'' 
of Theorem~\ref{na-qwm}. That is, $w_Q$  achieves the lower bound
$L_Q^{na}(F)$ defined in this theorem. 
Let $s_Q$ be the corresponding optimal choice for $s \in S'$.
We need to design a weight function $w_P$ 
which will show that $L_P(F) \geq L_Q^{na}(F)$.
One can simply define $w_P$ by: $w_P(x,y)=w_Q(x,y)$ if $F(x)=s_Q$ or
$F(y)=s_Q$;  $w_P(x,y)=0$ otherwise. Indeed, for any $i$ and any 
pair $(x,y)$ such 
that $w_P(x,y)>0$ we have $F(x)=s_Q$ or $F(y)=s_Q$, so that
$\max(wt(x)/v(x,i),wt(y)/v(y,i)) \geq L_Q^{na}(F)$.
 \hfill $\boxempty$\end{proof}
The nonadaptive quantum lower bound from Theorem~\ref{na-qwm} is therefore 
rather closely connected to adaptive probabilistic lower bounds: it is 
sandwiched between the weighted lower bound of Theorem~\ref{pwm} and
its unweighted $\max(m/l,m'/l')$ version.
Proposition~\ref{comparison} also implies that Theorem~\ref{na-qwm} can at best
prove an $\Omega(\log N)$ lower bound on the nonadaptive quantum complexity 
of Ordered Search. Indeed, by binary search the adaptive probabilistic 
complexity of this problem is $O(\log N)$.
In section~\ref{dual} we shall see that there is in fact a $\Omega(N)$ lower bound on the nonadaptive quantum complexity of this problem.
\begin{remark}
The connection between nonadaptive quantum complexity and 
adaptive probabilistic complexity that we have pointed 
out in the paragraph above is only a connection between 
the {\em lower bounds} on these quantities.
Indeed, there are problems with a high probabilistic query complexity
and a low nonadaptive quantum query complexity 
(for instance, Simon's problem~\cite{S94,KNPproba}).
Conversely, there are problems with a low probabilistic query complexity
and a high nonadaptive quantum query complexity (for instance, Ordered Search).
\end{remark}

\subsection{Proof of Theorem~\ref{na-qwm}}

As mentioned  in the introduction, we will treat the tuple $(i_1,\ldots,i_k)$
of queries made by a nonadaptive algorithm as a single ``super query'' made by 
an ordinary quantum algorithm (incidentally, this method could be used
to obtain lower bounds on quantum algorithm that make several rounds 
of parallel queries as in~\cite{GJ04}).
This motivates the following definition.
\begin{definition}
Let $\Sigma$, $\Gamma$ and $S$ be as in section~\ref{model}. Given an integer 
$k \geq 2$, we define:
\begin{itemize}
\item $^k\Sigma=\Sigma^k$, $^k\Gamma=\Gamma^k$ and 
$^k{S}=\left(
\Sigma^k \right) ^{\Gamma^k}$.
\item To the black box $x\in S$ we associate the ``super box'' 
 $^k{x} \in {}^k{S}$ such that 
if $I=(i_1, \dots, i_k)\in \Gamma^k$ then
$^k{x}(I)=(x(i_1), \dots, x(i_k))$.
\item $^k{F}(^k{x})=F(x)$.
\item If $w$ is a weight function for $F$ we define a weight function
$W$ for $^k{F}$ by $W(^k{x},^k{y})=w(x,y)$.
\end{itemize}
\end{definition}
Assume for instance that $\Sigma=\{0;1\}$, $\Gamma=[3]$, $k=2$, and 
that $x$ is defined
by: $x(1)=0$, $x(2)=1$ and $x(3)=0$. Then we have $^2{x}(1,1)=(0,0)$,
$^2{x}(1,2)=(0,1)$, $^2{x}(1,3)=(0,0)$ \dots


\begin{lemma}
\label{valid}
If $w$ is a valid weight function for $F$ then $W$ is a valid weight
function for $^kF$ and the minimal number
 of queries of a quantum algorithm computing
$^k{F}$ with error probability $\varepsilon$ satisfies:
$$Q_{2, \varepsilon}(^k{F}) \geq  C_{\varepsilon} \cdot
\min_{\substack{^k{x}, ^k{y}, I\\ W(^k{x}, ^k{y})> 0\\ ^k{x}(I)\neq ^k{y}(I)}}
\sqrt{
\frac{WT(^k{x})WT(^k{y})}{V(^k{x},I)V(^k{y},I)}
}.$$
\end{lemma}
\begin{proof}
Every pair $(x,y)\in S^2$ is assigned a non-negative weight
$W(^k{x},^k{y})= W(^k{y},^k{x})=w(x,y)=w(y,x)$ that satisfies
$W(^k{x},^k{y})=0$ whenever $F(x)=F(y)$. 
Thus we can apply Theorem~\ref{qwm} and we
obtain the announced lower bound.
\hfill $\boxempty$ \end{proof}

\begin{lemma}\label{facteurk}
Let $x$ be a black-box and $w$ a weight function. 
For any integer $k$ and any tuple $I=(i_1, \dots, i_k)$ we have

$$\frac{WT(^k{x})}{V(^k{x},I)} \geq
\frac{1}{k} \ \min_{j\in[k]} \  \frac{wt(x)}{v(x,i_j)}.$$
\end{lemma}
\begin{proof}
Let $m= \min_{j\in[k]} \  \frac{wt(x)}{v(x,i_j)}$. We have
$WT(^k{x})=wt(x)$ and: 

$$\begin{aligned}
V(^k{x},I) &= \sum_{^k{y}: ^k{x}(i)\neq ^k{y}(i)} W(^k{x},
^k{y})\\
&\leq \sum_{y: x(i_1)\neq y(i_1)} w(x,y) + \dots +
\sum_{y: x(i_k)\neq y(i_k)} w(x,y) \\
&= v(x,i_1) + \dots + v(x,i_k) \leq \ k \max_{j\in[k]}
v(x,i_j).\ \ \ \ \ \ \ \ \boxempty \end{aligned} $$
\end{proof}

\begin{lemma}\label{weak-qna-bound}
If $w$ is a valid weight function:

$$Q^{na}_{2,\varepsilon}(F) \geq C_{\varepsilon}^2
\min_{\substack{x,y\\ F(x)\neq F(y)}} \max
\left( 
\min_i \frac{wt(x)}{v(x,i)}, \min_i \frac{wt(y)}{v(y,i)}
\right).$$ 
\end{lemma}

\begin{proof}
Let $w$ be an arbitrary valid weight function and $k$ be an integer
such that  
$$k < C_{\varepsilon}^2
\min_{\substack{x,y\\ F(x)\neq F(y)}} \max
\left( 
\min_i \frac{wt(x)}{v(x,i)}, \min_i \frac{wt(y)}{v(y,i)}
\right).$$
We show that an algorithm computing $^kF$ with probability of error $\leq \varepsilon$ must make strictly more one than query to the ``super box'' $^kx$.
 This will prove
that for every such $k$ we have $Q^{na}_{2,\varepsilon}(F)>k$ and thus
our result. 

For every $x$ and $I$ we have 
$$\frac{WT(^k{x})}{V(^k{x},I)} \geq 1$$

and thus by lemma~\ref{facteurk} for every $x$, $y$ and $I=(i_1,
\dots, i_k)$: 

$$\begin{aligned}
\frac{WT(^k{x})}{V(^k{x},I)} \frac{WT(^k{y})}{V(^k{x},I)} &=
\min \left( \frac{WT(^k{x})}{V(^k{x},I)}, \frac{WT(^k{y})}{V(^k{x},I)}
\right) \max \left( \frac{WT(^k{x})}{V(^k{x},I)},
\frac{WT(^k{y})}{V(^k{x},I)} \right)\\
&\geq 
\max \left( \frac{WT(^k{x})}{V(^k{x},I)}, \frac{WT(^k{y})}{V(^k{x},I)}
\right) \\
&\geq \frac{1}{k} \max \left( \min_{j\in [k]} \frac{wt(x)}{v(x,i_j)},
\min_{l\in [k]} \frac{wt(y)}{v(x,i_l)} \right).
\end{aligned}$$

In order to apply Lemma~\ref{valid} we observe that:
$$\begin{aligned}
\min_{\substack{^k{x}, ^k{y}, I\\ W(^k{x}, ^k{y})> 0\\ ^k{x}(I)\neq
^k{y}(I)}}  \frac{WT(^k{x})WT(^k{y})}{V(^k{x},I)V(^k{y},I)} 
&\geq  \frac{1}{k} \min_{\substack{x, y, i_1, \dots, i_k\\ w(x,y)> 0\\
\exists m \ x(i_m)\neq y(i_m)}} \max \left( \min_{j\in [k]}
\frac{wt(x)}{v(x,i_j)}, \min_{l\in [k]} \frac{wt(y)}{v(x,i_l)}
 \right)  \\ 
&\geq  \frac{1}{k} \min_{\substack{x, y\\ F(x)\neq F(y)}} \max \left(
\min_i \frac{wt(x)}{v(x,i)}, \min_i \frac{wt(y)}{v(x,i)} \right)
\end{aligned}$$
By hypothesis on $k$, this expression is greater than $1/C_{\varepsilon}^2$.
Thus according to Lemma~\ref{valid} we have 
$Q_{2,\varepsilon}(^k{F})>1$, and $Q^{na}_{2,\varepsilon}(F)> k$.
\hfill $\boxempty$\end{proof}
We can now complete the proof of Theorem~\ref{na-qwm}.
Suppose without loss of generality that $F(S)=[m]$ 
and define for every $l\in [m]$:
$$a_l=C_{\varepsilon}^2 \min_{\substack{x,i\\ F(x)=l}}
\frac{wt(x)}{v(x,i)}.$$
 
Suppose also  without loss of generality that $a_1  \leq \dots \leq a_m$.
It follows immediately from the definition that
$$a_2= C_{\varepsilon}^2 \min_{\substack{x, y\\ F(x)\neq F(y)}} \max \left(
\min_i \frac{wt(x)}{v(x,i)}, \min_i \frac{wt(y)}{v(x,i)} \right),$$ 
and $$a_m= C_{\varepsilon}^2 \max_{l\in F(S)} \min_{\substack{x,i\\ F(x)=l}} 
\frac{wt(x)}{v(x,i)}.$$
By Lemma~\ref{weak-qna-bound} we have 
$Q^{na}_{2,\varepsilon}(F) \geq a_2$, but we would like to show that
$Q^{na}_{2,\varepsilon}(F) \geq a_m$. 
We proceed by reduction from the case when there are only two classes 
(i.e., $m=2$).
Let $G$ be defined by $$G(1)= \dots = G(m-1)=1$$ and $G(m)=m$. Applying
Lemma~\ref{weak-qna-bound} to $G o F$, we obtain that $Q^{na}_{2,\varepsilon}(G
o F) \geq  a_m$. But because the function $G o F$ is
obviously easier to compute than~$F$, we have 
$Q^{na}_{2,\varepsilon}(F) \geq Q^{na}_{2,\varepsilon}(G o F)$ and 
thus $Q^{na}_{2,\varepsilon}(F) \geq a_m$ as desired.

\section{From the Dual to the Primal}
\label{dual}

Our starting point in this section is the minimax method of Laplante and Magniez~\cite{LM04,SS05} as stated in \cite{HS05}:
\begin{thm}\label{thmadaptiveminmax}
Let $p: S\times \Sigma\rightarrow\mathbb{R}^+$ be the set of $|S|$
probability distributions such that $p_x(i)$ is the average
probability of querying $i$ on input $x$, where the average is taken
over the whole computation of an algorithm $\mathcal{A}$. Then the
query complexity of $\mathcal{A}$ is greater or equal to:
$$ \label{eqnadaptiveminmaxequaltospectral}
C_{\varepsilon} \max_{\begin{subarray}
\ \ \ \ x,y\\F(x)\neq
F(y)\end{subarray}}\frac{1}{\sum\limits_{\begin{subarray} \ \ \ \ \ i \\
x(i)\neq y(i)\end{subarray}} \sqrt{p_x(i)p_y(i)}}.$$
\end{thm}
Theorem~\ref{thmadaptiveminmax} 
is the basis for the following lower bound theorem.
It can be shown that up to constant factors, the lower bound given by Theorem~\ref{thmnonadaptiveminmaxqueryonebit} is always as good as the lower bound 
given by Theorem~\ref{na-qwm}.
\begin{thm}[nonadaptive quantum lower bound, primal-dual method]
\label{thmnonadaptiveminmaxqueryonebit}
Let $F:S \rightarrow S'$ be a partial function, where as usual 
$S=\Sigma^{\Gamma}$ is the set of black-box functions.
Let $$DL(F)= \min_p
\max\limits_{\begin{subarray}\ 
\ \ \ \ x,y\\ F(x)\neq
F(y)\end{subarray}}\frac{1}{\sum\limits_{\begin{subarray} \ \ \ \ \ i
\\ x(i)\neq y(i)\end{subarray}} p(i)}$$
and $$PL(F)=\max_w \frac{\sum\limits_{x,y}w(x,y)}{\max\limits_i
\sum\limits_{\begin{subarray}\  \ x,y\\x_i\neq y_i\end{subarray}}w(x,y)}$$
where the $\min$ in the first formula is taken over all
probability distributions $p$ over $\Gamma$, and the $\max$ in the second
formula is taken over all valid weight functions $w$.
Then $DL(F)=PL(F)$ and
we have the following nonadaptive query complexity lower bound:
$$Q_{2,\varepsilon}(F)\geq C_{\varepsilon}DL(F)=C_{\varepsilon}PL(F).$$
\end{thm}
\begin{proof} 
We first show that $Q_{2,\varepsilon}(F)\geq C_{\varepsilon}DL(F)$.
Let $\mathcal {A}$ be a
nonadaptive quantum algorithm for $F$. 
Since $\mathcal {A}$ is nonadaptive, the probability $p_x(i)$ 
of querying $i$ on input $x$ is independent of $x$. We denote it by $p(i)$.
Theorem~\ref{thmadaptiveminmax} shows that the query complexity of 
$\mathcal {A}$ is greater or equal to 
$$C_{\varepsilon} \max\limits_{\begin{subarray}\ 
\ \ \ \ x,y\\ F(x)\neq
F(y)\end{subarray}}\frac{1}{\sum\limits_{\begin{subarray} \ \ \ \ \ i \\
x(i)\neq y(i)\end{subarray}}p(i)}.$$
The lower bound $Q_{2,\varepsilon}(F)\geq C_{\varepsilon}DL(F)$ 
follows by minimizing over $p$.

It remains to show that $DL(F)=PL(F)$.
Let
$$L(F)=\min_p\max\limits_{\begin{subarray}\ \ \ \ \ x,y\\
F(x)\neq F(y)\end{subarray}}\sum\limits_{\begin{subarray} \ \ \ \ \ i \\
x(i)= y(i)\end{subarray}}p(i).$$
We observe that 
$L(F)$ is the optimal solution of the following linear program:
minimize $\mu$ subject to the constraints
\begin{eqnarray}\label{eqnlponebit}
\forall x,y \mbox{ \rm such that } f(x)\neq f(y):\
\mu-\sum\limits_{\begin{subarray} \ \ \ \ \ i \\ 
x(i)\neq y(i)\end{subarray}}p(i)\geq0, \nonumber\\
\mbox{ \rm and to the constraints }\sum_{i=1}^N p(i)=1 
\mbox{ \rm and } \forall i\in[N]: \ p(i)\geq0\nonumber .
\end{eqnarray}
Clearly, its solution set is nonempty. 
Thus $L(f)$ is the optimal solution of the dual linear program:
maximize $\nu$ subject to the constraints
\begin{eqnarray}\label{eqndponebit}
\forall i\in [N]:\ \nu-\sum_{\begin{subarray}\ \ \ x,y\\x_i=y_i\end{subarray}}w(x,y)\leq0\nonumber \\
\forall x,y:\ w(x,y)\geq0, \text{ and }  w(x,y)=0\ \text{if $F(x)=F(y)$}\nonumber\\
\text{and to the constraint} \sum_{x,y}w(x,y)=1.\nonumber
\end{eqnarray}
Hence
$\displaystyle L(F)=\max_w \min_i
\frac{\sum\limits_{x_i=y_i}w(x,y)}{\sum\limits_{x,y}w(x,y)}$ 
and  $DL(F)= \frac{1}{1-L(F)}=PL(F).$
\hfill $\boxempty$ \end{proof}

\subsection{Application to Ordered Search and Connectivity}

\begin{prop}\label{propconclusionfororderedsearching}
For any error bound $\varepsilon\in[0,\frac{1}{2})$ 
we have $$Q_{2, \varepsilon}^{na}(\text{Ordered Search})\geq
C_{\varepsilon}(N-1).$$ 
\end{prop}
\begin{proof}
Consider the weight function
$\displaystyle w(x,y)=
\begin{cases}
1 & \text{if $|F(y)-F(x)|=1$}, \\ 0 &\text{otherwise}.
\end{cases}$
Thus $w(x,y)=1$ when the leftmost 1's in $x$ and $y$ are adjacent.
Hence $\sum\limits_{x,y}w(x,y)=2(N-2)+2$. Moreover, if $w(x,y) \neq 0$ and
$x_i \neq y_i$ then $\{F(x),F(y)\} = \{i,i+1\}$.
Therefore, $\max\limits_i\sum\limits_{\begin{subarray}\  \ x,y\\x_i\neq y_i\end{subarray}}w(x,y)=2$ and the result follows from
Theorem~\ref{thmnonadaptiveminmaxqueryonebit}.
\hfill $\boxempty$\end{proof}

Our second application of Theorem~\ref{thmnonadaptiveminmaxqueryonebit}
is to the graph connectivity problem. We consider the adjacency matrix model:
$x(i,j)=1$ if $ij$ is an edge of the graph. 
We consider undirected, loopless graph
so that we can assume $j<i$. For a graph on $n$ vertices, 
the black box $x$ therefore has 
$N=n(n-1)/2$ entries. We denote by $G_x$ the graph represented by $x$.
\begin{thm}\label{thmgraphproblem}
For any error bound $\varepsilon\in[0,\frac{1}{2})$, we have
$$Q_{2,\varepsilon}^{na}(\text{Connectivity})\geq C_{\varepsilon}n(n-1)/8.$$
\end{thm}

\begin{proof}
We shall use essentially the same weight function as in~(\cite{DHHM06},
Theorem~8.3).
Let $X$ be the set of all adjacency matrices of a
unique cycle, and $Y$ the set of all adjacency matrices with exactly
two (disjoint) cycles.
For $x \in X$ and $y \in Y$,  we set $w(x,y)=1$ 
if there exist 4 vertices $a,b,c,d\in[n]$ 
such that the only differences
between $G_x$ and $G_y$ are that:
\begin{enumerate}
\item  $ab, cd$ are edges in $G_x$ but not
in $G_y$.
\item $ac, bd$ are edges in $G_y$ but not in $G_x$. 
\end{enumerate}
We claim that
\begin{equation} \label{connect}
\max_{ij}\sum\limits_{\begin{subarray}\
\ x \in X,y \in Y\\x(i,j)\neq
y(i,j)\end{subarray}}w(x,y)=\frac{8}{n(n-1)}\sum\limits_{\begin{subarray}\ \
x \in X,y \in Y\\x(i,j)\neq y(i,j)\end{subarray}}w(x,y).
\end{equation}
The conclusion of Theorem~\ref{thmgraphproblem} will then follow directly from
Theorem~\ref{thmnonadaptiveminmaxqueryonebit}.
By symmetry, the function that we are maximizing on the left-hand side 
of~(\ref{connect}) is in fact independent of the edge $ij$.
We can therefore replace the max over $ij$ by an average over~$ij$:
the left-hand side is equal to
$$\frac{1}{N}\sum\limits_{\begin{subarray}\
\ x \in X,y \in Y\\\end{subarray}}w(x,y)|\{ij;\ x(i,j)\neq y(i,j)\}|.$$
Now, the condition $x(i,j)\neq y(i,j)$ holds true if and only if $ij$
is one of the 4 edges $ab$, $cd$, $ac$, $bd$ defined at the beginning 
of the proof. 
This finishes the proof of~(\ref{connect}), 
and of Theorem~\ref{thmgraphproblem}.\hfill $\boxempty$
\end{proof}
A similar argument can be used to show that testing whether a graph is 
bipartite also requires $\Omega(n^2)$ queries.

\section{Some Open Problems}

For the ``1-to-1 versus 2-to-1'' problem, one
would expect a higher quantum query complexity in the nonadaptive setting
than in the adaptive setting. This may be difficult to establish since
the adaptive lower bound~\cite{AS04} is based on the polynomial method.
Hidden Translation~\cite{FIMSS03} 
(a problem closely connected to the dihedral hidden subgroup problem) 
is another problem of interest. No lower bound is known in the 
adaptive setting, so it would be natural to look first for a nonadaptive 
lower bound.
Finally, one would like to identify some classes of problems for which
adaptivity does not help quantum algorithms.

{\small {\bf Acknowledgements:} 
This work has benefited from discussions with Sophie Laplante, Troy Lee, 
Frédéric Magniez and Vincent Nesme.

Email addresses: \url{[Pascal.Koiran,Natacha.Portier]@ens-lyon.fr},\\
\url{juergen_landes@yahoo.de}, \url{phyao1985@gmail.com}.

\bibliographystyle{plain}
\bibliography{quantum}

\end{document}